\documentclass{elsarticle}

\usepackage{hyperref}
\usepackage{amsmath}
\usepackage{amssymb}
\usepackage{amsthm}
\usepackage{amsfonts}
\newtheorem{theorem}{Theorem}
\newtheorem{proposition}[theorem]{Proposition}

\def\ri{\mathrm i}
\def\e{\mathrm e}
\def\C{\mathbb C}
\def\R{\mathbb R}
\newcommand{\be}{\begin{equation}}
\newcommand{\ee}{\end{equation}}

\begin{document}

\begin{frontmatter}

\title{Linear Instability of the Peregrine Breather: Numerical and Analytical Investigations}

\author[mymainaddress]{A. Calini}
\address[mymainaddress]{College of Charleston, SC}

\author[mysecondaryaddress]{C. M. Schober\corref{mycorrespondingauthor}}
\address[mysecondaryaddress]{University of Central Florida, FL}
\cortext[mycorrespondingauthor]{Corresponding author}
\ead{cschober@ucf.edu}

\author[mysecondaryaddress]{M. Strawn}

\begin{abstract}
We study the linear stability of the Peregrine breather both numerically and with analytical arguments  based on its derivation as the singular limit of a single-mode spatially periodic breather as the spatial period becomes infinite.
By constructing solutions of the linearization of the nonlinear Schr\"odinger equation in terms of quadratic products of components of the eigenfunctions of the Zakharov-Shabat system, we show that the Peregrine breather is linearly unstable. 
A numerical study employing a highly accurate Chebychev pseudo-spectral integrator confirms exponential growth of random initial perturbations of the Peregrine breather.

\end{abstract}

\begin{keyword}
Rogue wave\sep Peregrine solution\sep Stability
\end{keyword}

\end{frontmatter}

%\linenumbers

\section{Introduction}

The Peregrine breather \cite{Peregrine1983}:
\begin{equation}
\label{Peregrine}
u_P(x,t) = a \e^{2\ri a^2t}\left(-1+  \frac{ 16\ri a^2t+4 }{4a^2x^2+16a^4t^2+1}\right), \quad a\in \R^+ ,
\end{equation}
the simplest amongst the rational solutions of the focusing nonlinear Schr\"odinger (NLS) equation
\begin{equation}
\ri u_{t} + u_{xx}  + 2|u|^2u = 0,
\label{nls}
\end{equation}
has often been described as a rogue wave model \cite{aat09,as-ca09,Chabchoub2013,aat09} due to its localized shape in both space and time,  rising above a uniform background modeled by the Stokes wave solution 
\begin{equation}
\label{stokes}
u_a(t)=\e^{2\ri a^2 t}.
\end{equation}

The NLS equation \eqref{nls} describes the dynamics of deep-water waves; within such a framework rogue waves  develop due to a combination of modulational instability (MI) of the underlying background state and nonlinear focusing. 
Under periodic boundary conditions on the spatial interval $[0,L]$, the onset and details of the MI depend on the amplitude of the background and the size $L$ of the domain. In the case of the Stokes wave \eqref{stokes}, for fixed $a$, there is a threshold $L_0$ for which MI does not occur. The larger the period $L>L_0$, the more unstable modes emerge as the MI develops. This can easily be seen by considering a small perturbation of the form 
$u(x,t) = u_a(t)[1 + \epsilon(x,t)]$, $|\epsilon| <<1$, where $\epsilon$ satisfies the linearized NLS equation
\begin{equation}
\ri\epsilon_{t} + \epsilon_{xx}  + 2a^2(\epsilon + \epsilon^*) = 0.
\label{LNLS}
\end{equation}
Representing $\epsilon$ as a Fourier series with modes  $\epsilon_j \sim \e^{\ri\mu_j x + \sigma_j t}$, $\mu_j = 2\pi j/L$, gives 
$\sigma_j^2 = \mu_j^2\left(4a^2 - \mu_j^2 \right)$. It follows that the MI develops when $L>\pi/a$; moreover, letting $N=\lfloor {aL}/{\pi} \rfloor$ (the minimum integer greater than $aL/\pi$), $u_a$ is unstable with $N$ unstable modes.

Linear instabilities of more general spatially periodic NLS solutions can be characterized in terms of the Floquet spectrum of the linear operator  $\mathcal{L}$ in the Zakharov-Shabat (Z-S) system \cite{AKNS}:
\begin{equation}
\label{zs}
\begin{split}
\mathcal{L}(u) \Phi& = \lambda \Phi,\qquad\mathcal{L}(u)=\begin{pmatrix} \ri{\partial_x} & u \\ -u^* & -\ri{\partial_x} \end{pmatrix},
 \\
  \Phi_t & = \begin{pmatrix} -2\ri \lambda^2 +\ri |u|^2 & 2\ri \lambda u -u_x \\ 2 \ri \lambda u^* +u^*_x & 2\ri \lambda^2 -\ri |u|^2 \end{pmatrix} \Phi,
\end{split}
\end{equation}
whose compatibility condition is the NLS equation \eqref{nls}. 

Given an NLS solution satisfying $u(x+L, t)=u(x, t)$, we
say that $\omega$ is a {\em Floquet exponent} for a given value of $\lambda$
if there is a nontrivial solution of \eqref{zs} satisfying $\Phi(x+L,t)=e^{\ri\omega}\Phi(x,t)$.
Then the discrete and continuous {\em Floquet spectra} of $u$ are defined as follows:
$$
%\begin{eqnarray}
%\nonumber
\sigma_d(u)=\{\lambda  \in \C\, | \, e^{i\omega} =\pm 1\},
%\exists \,\vphi\ne0 \text{ satisfying \eqref{AK}  with }\vphi(x+L, t) = \pm \vphi(x,t) \},
\qquad
%\nonumber
\sigma_c(u) =\{\lambda \in \C\, | \, \omega \in \R\}.
%\exists \,\vphi\ne0 \text{ satisfying \eqref{AK}  with } \vphi(x,t) \text{ uniformly bounded for all }x\in \R\}.
%\end{eqnarray}
$$
Both spectra are symmetric under complex conjugation. The continuous spectrum includes the entire real axis and (typically) complex bands (as $\mathcal{L}$ is not self-adjoint). Given a fundamental solution matrix 
$M(x, t;\lambda)$ of~\eqref{zs}, the \emph{Floquet discriminant} is defined by
\[
\Delta(\lambda) = \mbox{trace} \,[M(x,t)^{-1} M(x+L,t) ]= 2\cos(\omega),
\]
which is an analytic function of $\lambda$.
Then, points of $\sigma_d$ are classified
as simple or multiple {\em periodic points} according to the order of vanishing of $\Delta^2-4$.
In addition, {\em critical points} of $\sigma_c$ are those where $d\omega/d\lambda=0$ for an analytic
branch of $\omega(\lambda)$.

\begin{figure}[!ht]
\centerline{
\includegraphics[width=.4\textwidth]{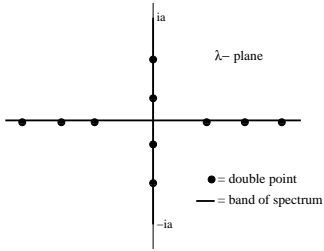}
}
\caption{Floquet spectrum of the Stokes wave $u_a$ with $\displaystyle \lfloor {aL}/{\pi} \rfloor=2.$}
\label{floquet_spc}
\end{figure}

The Floquet discriminant for the Stokes wave is  $\Delta(\lambda)= 2 \cos(\sqrt{a^2 + \lambda^2} L)$ and its Floquet spectrum has just one band of complex spectrum $[-\ri a, \ri a]$,  and infinitely many double points 
\begin{equation}
\nu_j^2 = \left(\frac{j\pi}{L}\right)^2 - a^2,\quad j \ne 0,
\label{imdps}
\end{equation}
as illustrated in Figure \ref{floquet_spc}. 

Note that when $L>\pi/a$, $\lfloor {aL}/{\pi} \rfloor$  of the double points \eqref{imdps} are imaginary, reflecting the well-known correspondence between complex double points and linear unstable modes \cite{limc94}.

Given an unstable Stokes wave with $N$ unstable modes and a positive integer $M\leq N$, its $M$-dimensional heteroclinic orbits are localized in time, and referred to as the $M$-mode spatially periodic breathers (SPB) \cite{dt99,osonse00,as-ca09,cs02,Islas2004}.  In the context of modeling rogue waves, this class of solutions was discussed by the authors in a series of papers~\cite{cs02, cs09, cs12, cs13}, where their stability with respect to small changes in initial conditions is proposed as the main criterion for determining which, among such solutions, are observable and reproducible in an experimental setting. 
A broad numerical and analytical study established that general SPBs are linearly unstable \cite{cs2015}, with the exception of the  \emph{maximal} SPBs (i.e. those with  $N$ modes over a Stokes wave with $N$ 
unstable modes). The maximal SPBs are neutrally stable and are thus good candidates for modeling realistic (observable and reproducible) rogue waves \cite{cs13}.

A convenient way to construct SPBs is by means of the well-known gauge formulation of the B\"acklund transformation, originally due to \cite{sazu87}. Its main advantage for our purposes is the simultaneous derivation of the SPB as well as  its associated eigenfunctions. One can then construct \emph{squared eigenfunction} solutions of the linearized NLS~\eqref{LNLS} in terms of quadratic products of eigenfunction components and determine the linear stability type of the SPB from their behavior over time.

In this work, we propose a similar framework for studying the linear stability of the Peregrine breather~\eqref{Peregrine}, derived as the singular limit of a single-mode SPB as the spatial period becomes infinite.  The large $L$ limit can be computed at all stages of the construction outlined above, leading to explicit expressions for the Peregrine breather, its associated eigenfunctions, and the resulting family of squared eigenfunction solutions of the linearized NLS, which carry information on the linear stability of the limiting solution. Section~\ref{GBT} briefly describes the gauge-B\"acklund transformation for the NLS equation and the construction of the single-mode SPB. Section~\ref{PBlim} derives the infinite-period limit of the SPB and the eigenfunctions of the Peregrine breather. In Section 4 the linear stability of the single-mode SPB and of the Peregrine breather are discussed.  Section 5 describes the numerical study confirming the instability of the Peregrine breather. This approach can be easily, though tediously, extended to the higher-order rational solutions of the NLS to show that any of these solutions is also linearly unstable.

\section{Constructing the SPB via Gauge-B\"acklund Transformation}
\label{GBT}

Let  $u$ be an NLS solution, and for a fixed value $\nu$ of the spectral parameter, let 
 $\Phi=(\phi_1, \phi_2)^T$ be a non-trivial solution of the Z-S system at $(u,\nu)$. Then
  \begin{equation}
\label{back}
u_h(x,t;\nu) = u - 2(\nu - \nu^*) 
\frac{\phi_1 \phi^*_2}{|\phi_1|^2 + |\phi_2|^2}
\end{equation}
is also a solution of the NLS equation, known as the {\em B\"acklund transformation} of $u$ based at $\nu$. Note that $\nu$ must be non-real for the transformation to be non-trivial.

Moreover, if $\Phi$ solves the Z-S system at $(u,\lambda)$ for $\lambda\not=\nu,\, \nu^*$, then 
\begin{equation}
\label{eigen}
\Phi_h (x,t; \lambda, \nu) = G(x,t; \lambda, \nu) \Phi (x,t; \lambda)
\end{equation}
solves the Z-S system at $(u_h, \lambda)$, where the \emph{gauge matrix} $G$ is given by
 \begin{equation}
G= N \left( \begin{array}{cc} 
-\lambda + \nu & 0 \\ 0 & -\lambda + \nu^*
\end{array} \right) N^{-1}, \qquad
N= \left( \begin{array}{cc} \phi_1 & - \phi^*_2 \\
\phi_2 & \phi^*_1 \end{array} \right).
\end{equation}

If $u$ is a spatially periodic NLS potential with complex double points in its Floquet spectrum, and $\nu$ is one such point, formula~\eqref{back} yields a heteroclinic orbit of $u$, and iterated B\"acklund transformations will lead to an explicit representation of the stable/unstable manifold of $u$. \smallskip

The simplest non-trivial example takes the `seed potential' to be an unstable Stokes wave $u=u_a(t):=a \e^{2\ri a^2 t}$ with $N$ imaginary double points $\{\nu_j\}$ given by~\eqref{imdps}, and constructs a non-trivial solution of the associated Z-S system as a complex linear combination of its Bloch eigenfunctions
\begin{align}
\Phi^\pm(x, t;\lambda) &= \frac{\ri}{2k}\e^{\mp \ri\frac{\pi}{4}}\e^{\pm \ri\frac{p}{2}}\left[\begin{array}{c}
a \e^{\ri a^2 t}\\\pm a \e^{\mp \ri p}\e^{-\ri a^2t}
\end{array}  \right]\e^{\pm \ri (kx+2k\lambda t)}\label{efnsPlanewave},
\end{align}
where $k^2 -\lambda^2 = a^2$ and $p$ is defined by $k\pm\lambda = a \e^{\mp \ri p}.$

Given $c_\pm$, an arbitrary pair of complex numbers, and setting $c_+/c_- = \e^{\rho+\ri\beta}$, one constructs
\begin{align*}
&\Phi\left(x,t;\lambda\right) = c_+\phi^+\left(x, t;\lambda\right)+c_-\phi^-\left(x, t;\lambda\right)\\
&= \frac{\ri ac_-}{2k}\left[\begin{array}{c}
\left( \e^{\rho} \e^{\ri\beta} \e^{-\ri\pi/4}\e^{\ri p/2}\e^{ \ri \left(kx+2k\lambda t\right)}+ 		\e^{\ri\pi/4}\e^{-\ri p/2}\e^{- \ri \left(kx+2k\lambda t\right)}\right)\e^{\ri a^2t}\\
\left(  \e^{\rho} \e^{\ri\beta} \e^{-\ri\pi/4}\e^{-\ri p/2}\e^{ \ri \left(kx+2k\lambda t\right)} -		\e^{\ri\pi/4}\e^{\ri p/2}\e^{- \ri \left(kx+2k\lambda t\right)}\right)\e^{-\ri a^2t}
\end{array}  \right].
\end{align*}
Evaluating $\Phi$ at $\nu=\nu_j=\ri \alpha_j$, $\alpha_j =\sqrt{(j\pi/L)^2-a^2}$, one computes the gauge matrix
\begin{equation}
G(x, t; \lambda, \alpha_j) 
=\left[ \begin{array}{cc}
-\lambda+\ri\alpha_j A_j
&\ri\alpha_j B_j\\
 \ri\alpha_j B_j^*
  &-\lambda-\ri\alpha_j A_j\\
\end{array}\right], \label{GaugeAB}
\end{equation}
 where 
\begin{align}
A_j &= \frac{|\phi_1|^2-|\phi_2|^2 }{|\phi_1|^2+|\phi_2|^2 }
	=\frac{\cos p_j\sin{\left(2k_jx+\beta\right)}}{\cosh{\left(\rho-\sigma_j t\right)}+\sin p_j\cos\left(2k_jx+\beta\right)}\label{Aperiodic},\\
B_j &= \frac{2\phi_1\phi_2^*}{|\phi_1|^2+|\phi_2|^2 }\nonumber\\
&=\frac{\ri\sin p_j\cosh{\left(\rho-\sigma_j t\right)}+\cos p_j\sinh{\left(\rho-\sigma_j t\right)} 
	+\ri \cos{\left(2k_jx+\beta\right)}}{\cosh{\left(\rho-\sigma_j t\right)}+\sin p_j\cos\left(2k_jx+\beta\right)}\e^{2\ri a^2 t},\label{Bperiodic}
\end{align}
and $k_j =k(\nu_j)= \pi j/L, \, \sigma_j = 4 k_j\alpha_j, \, \cos p_j = k_j/a, \, \sin p_j=-\alpha_j/a$.

Then, letting $\tau_j = \rho - \sigma_j t$, the B\"acklund transformation of $u_a$ based at $\nu_j$ is given by the following:
\begin{align}
&u_h(x,t;\nu_j) = u_a(x,t) - 2\ri\alpha_j\, B_j=
u_a(x,t) + 2\ri a\sin p_j\, B_j\nonumber\\
&= a\e^{2\ri a^2t} + 2\ri a\sin p_j\e^{2\ri a^2t}\frac{\cos p_j\sinh\,\tau_j+\ri\sin p_j\cosh\,\tau_j+\ri\cos\left(2k_jx+\beta\right)}{\cosh\,\tau_j+\sin p_j\cos\left(2k_jx+\beta\right)}\nonumber\\
&= a\e^{2\ri a^2t}\left( \frac{\ri\sin 2p_j\tanh\,\tau_j+\cos 2p_j-\sin p_j\cos\left(2k_jx+\beta\right)\mbox{ sech }\tau_j}{1+\sin p_j\cos\left(2k_jx+\beta\right)\mbox{ sech }\tau_j} \right).\label{HomoclinicSech}
\end{align}

\begin{figure}[!ht]
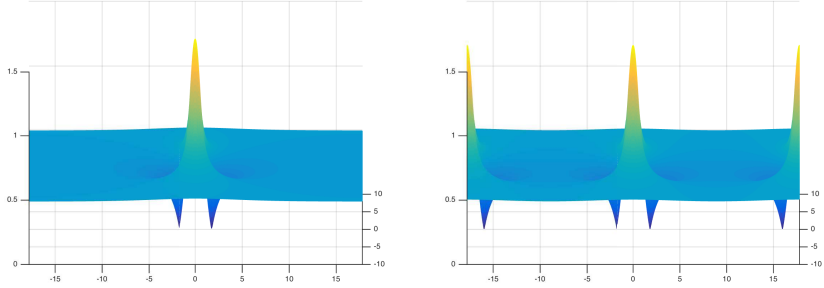

\centerline{
\includegraphics[width=.47\textwidth]{schober_fig2a}
\includegraphics[width=.47\textwidth]{schober_fig2b}
}
\caption{Amplitudes of two single mode SPBs over an unstable Stokes wave with six unstable modes ($L = 40, a = 0.5$).:
(left)  $|u_h(x,t;\nu_1)|$ and (right)  $|u_h(x,t;\nu_2)|$.}
\label{SPB_surfs}
\end{figure}

Formula \eqref{HomoclinicSech} describes the single-mode SPB with wave number $k_j$ \cite{Osborne2014,cs12}. See Figure~\ref{SPB_surfs} for the amplitude plots of two such solutions. Note that $\lim_{t\rightarrow \pm \infty} u_h=\e^{\pm 2 \ri p_j} u_a$, thus $u_h$ is the heteroclinic orbit of the Stokes wave associated with the $j$-th unstable mode. (This is also reflected in the temporal growth rate $\sigma_j$.) 

The Bloch eigenfunctions for $u_h(x,t;\nu_j)$ are obtained by means of the gauge transformation~\eqref{eigen} as 
$\Phi_h^\pm(x,t;\lambda, \nu_j) = G(x,t; \lambda, \nu_j)\Phi^\pm(x,t;\lambda).$  We return to them later when discussing the linear stability of the SPB.

\section{The Peregrine breather as limiting case of the Spatially Periodic Breather}
\label{PBlim}

The Peregrine breather was first derived from the Ma soliton letting its temporal period go to infinity \cite{Peregrine1983}. However, its expression can also be obtained as a singular limit of any of the single-mode SPBs constructed above, as the spatial period $L\rightarrow \infty$. 

To illustrate the derivation we let $\rho = \beta = 0$ (selecting the particular $\Phi$ for which $c_+=c_-$) and compute
\begin{equation}
\label{Bnoshift}
B_j =\frac{\phi_1 \phi^*_2}{|\phi_1|^2 + |\phi_2|^2}=\e^{2\ri a^2t}\,\frac{\ri\sin p_j-\cos p_j\tanh \sigma_jt+\ri\cos 2k_jx\mbox{ sech } \sigma_jt}{1+\sin p_j\cos 2k_jx \mbox{ sech }\sigma_j t}.
\end{equation}
For fixed $j$, as $L\to\infty$,  $k_j = \pi j/L\to 0$, $\nu_j\to \ri a$ and 
$\sigma_j \to 4k_ja$.

Substituting the small-$k_j$ perturbation expansions:
\begin{align*}
& \sin p_j\cos 2k_jx\mbox{ sech } \sigma_j t 
	= -1+\frac{k_j^2}{2a^2}\left(4a^2x^2+16a^4t^2+1 \right)+\cdots,\\
%%%%%%%%%%%%%%%%%%%%%%
&\tanh \sigma_j t =  4k_jat +\cdots,\\
%%%%%%%%%%%%%%%%%%%%%%
&\cos p_j\sin 2k_j x\mbox{ sech } 4k_ja t  
	= \frac{2k_j^2x}{a}+\cdots,
\end{align*}
in equation~\eqref{Bnoshift} gives\footnote{The subscript $P$ stands for ``Peregrine", for the limiting NLS solution.}

\begin{equation}
	{B_P}:= \lim_{k_j\to 0} B _j=  \e^{2\ri a^2t}\left(\frac{-8a^2t +2\ri}{4a^2x^2+16a^4t^2+1}-\ri\right), \label{Binfty}
\end{equation}
which substituted in~\eqref{HomoclinicSech} gives the well-known expression of the Peregrine breather:
\begin{equation}
u_P(x,t) = \lim_{k_j\to 0} u_h(x,t;\nu_j) = a \e^{2\ri a^2t}\left(-1+  \frac{ 16\ri a^2t+4 }{4a^2x^2+16a^4t^2+1}\right) .
\end{equation}
Note that $u_P$ decays polynomially in $x$ to the Stokes wave as $x\rightarrow \pm \infty$ and decays at an $O(t^{-1})$ rate to  phase shifts of the Stokes wave as $t\rightarrow \pm \infty$. The amplitude plot of $u_P$ is
 visually indistinguishable from that of the single mode SPB $u_h(x,t;\nu_1)$  on the interval $[-L/2, L/2]$ (where $L$ is the spatial period of the SPB) as shown in Figure~\ref{SPB_surfs}.

In a similar way, we obtain a fundamental set of eigenfunctions satisfying the Z-S system at $(u_P, \lambda)$.  
Expanding expression~\eqref{Aperiodic} for small $k_j$ as above, gives:
\begin{equation}
	A_P= \lim_{k_j\to 0} A_j = \frac{4ax}{4a^2x^2+16a^4t^2+1}\label{Ainfty}.
\end{equation}
Replacing $A_j$ and $B_j$ with their limiting expressions in the entries of the gauge matrix, we obtain 
\begin{align}
&\Phi_P^\pm (x,t;\lambda)= \frac{\ri}{2k}\e^{\mp \ri\frac{\pi}{4}}\e^{\pm \ri\frac{p}{2}} \left[ \begin{array}{cc}
- \lambda+\ri a A_P &\ri a B_P\\ \ri a B_P^*  &- \lambda-\ri a A_P\\
\end{array}\right]
\left[\begin{array}{c}
a\e^{\ri a^2t}\\\pm a \e^{\mp \ri p}\e^{-\ri a^2t}
\end{array}  \right]\nonumber\\
&\times\e^{\pm \ri (kx+2k\lambda t)}\nonumber\\
%%%%%%%%%%%%%
%%%%%%%%%%%%%
&= \frac{\ri a}{2k}\e^{\mp \ri\frac{\pi}{4}}\e^{\pm \ri\frac{p}{2}} \left[ \begin{array}{c}
\e^{\ri a^2t}\left(- \lambda + \frac{4\ri a^2x}{m} \pm \ri a \e^{\mp \ri p}\frac{-8a^2t +\ri\left(1 -4a^2x^2-16a^4t^2 \right)}{m}\right)\\
\e^{-\ri a^2t}\e^{\mp \ri p}\left(\mp \lambda  + \ri a\e^{\pm \ri p}\frac{-8a^2t -\ri\left(1 -4a^2x^2-16a^4t^2 \right)}{m}  \mp \frac{4\ri a^2x}{m}\right)\\
\end{array}\right]\nonumber\\
&\times\e^{\pm \ri (kx+2k\lambda t)},\label{pereigen}
\end{align}
where $m = 4a^2x^2+16a^4t^2+1$. Note that $\displaystyle \lim_{k_j\to 0} \text{det}\left(G(x,t;\lambda, \alpha_j)\right)=\lambda^2+a^2$, which is non-zero except at $\lambda=\pm \ri a$, thus in general $\Phi_P^\pm$ are linearly independent.

\section{Linear Stability: SPB versus Peregrine Breather}

To analyze the linear stability analysis of the SPB and its large-$L$ limit, the Peregrine breather, we make use of the following well-known fact:
\begin{proposition}
 Let $\Phi$ and $\Psi$ solve the Z-S system at $(u,\lambda)$. Then:
\begin{equation}
\label{SqeSol}
\epsilon_1(x,t) =\phi_1{\psi}_1 +\phi^*_2\psi^*_2, \qquad \text{and} \qquad 
\epsilon_2(x,t) =\ri( \phi_1{\psi}_1 -\phi^*_2\psi^*_2),
\end{equation}
\smallskip
solve the linearized NLS equation (\ref{LNLS}) at $u$.
\end{proposition}

\subsection*{Spatially Periodic Breather: saturation of instability}
The eigenfunctions of $u_h(x,t;\nu_j)$:
\begin{align}
&\Phi_h^\pm(x,t,\lambda;\nu_j) = G(x,t; \lambda; \nu_j)\Phi^\pm(x,t,\lambda)\nonumber\\
&=\frac{\ri a}{2k}\e^{\mp \ri\pi/4}\e^{\pm \ri p/2} \left[ \begin{array}{cc}
-\lambda+\nu_j A_j
&\nu_j B_j\\
 \nu_j B_j^*
  &-\lambda-\nu_j A_j\\
\end{array}\right]\left[\begin{array}{c}
\e^{\ri a^2t}\\\pm  \e^{\mp \ri p}\e^{-\ri a^2t}
\end{array}  \right]\e^{\pm \ri \left(kx+2k\lambda t\right)},\label{Phiperiodic}
\end{align}
will produce spatially periodic solutions $\epsilon_1, \epsilon_2$ of the linearized NLS at $u_h$  via~\eqref{SqeSol} only if $\Phi_h^\pm$ are periodic or antiperiodic functions of period $L$, i.e.~only if $\lambda=\nu_m=\ri \alpha_m$, one of the imaginary double point of the underlying Stokes wave. Moreover, because the entries of the gauge matrix are bounded functions of time $t$ (see \eqref{Aperiodic}, \eqref{Bperiodic}), the factor 
\[
\left.\e^{\pm \ri \left(kx+2k\lambda t\right)}\right|_{\lambda=\ri\alpha_k}=\e^{\pm \ri k_mx}\e^{\mp 2k_m \alpha_m t}
\]
is alone responsible for any exponential growth in time of $\epsilon_1, \epsilon_2$.

On the other hand,  $\mathrm{det}\left(G(x, t;\lambda; \nu_1\right))=\lambda^2 -\nu^2_j$ vanishes when $\lambda=\nu_j$, at which $\Phi_h^\pm(x,t,\lambda;\nu_j)$ become linearly dependent. One can show (see~\cite{cs12} for details) that $\Phi_h^\pm(x,t,\nu_j;\nu_j)$  is bounded in time, and that the $j$-th unstable mode of the Stokes wave has  {\em saturated} to a neutral mode for the associated SPB.

In conclusion, the SPB over a Stokes wave with one unstable mode is linearly stable, while each SPB constructed from a Stokes wave with $N$ unstable modes is itself unstable, with $N-1$ unstable modes.

\subsection*{The Peregrine Breather: linear instability}

A similar argument shows that the Peregrine Breather is linearly unstable. The associated eigenfunctions \eqref{pereigen} produce solutions $\epsilon_1$ and $\epsilon_2$ of the linearized NLS equation at $u_P$ that are bounded in $x$ if and only if $k(\lambda)=\sqrt{a^2+\lambda^2}\in \R$, i.e. $\lambda \in \R\cup [-\ri a, \ri a].$ In particular, when $\lambda=\ri \alpha$, with $|\alpha|<a$, then $\Phi_P^\pm$ are linearly independent and $\epsilon_{1,2}$ grow exponentially in time as $\displaystyle \e^{4|\alpha |\sqrt{a^2-\alpha^2}t}.$ Observe that the maximal growth rate is $\sigma_{\max} = 2a^2$, with the most unstable mode occurring when  $\displaystyle \lambda_m=\pm \ri \frac{\sqrt{2}}{2} a.$

\section{Comparison with Numerical Experiments}
\label{CPS4}

In this final section we present numerical evidence confirming the instability of the Peregrine breather.
We use a highly accurate  Chebyshev pseudo-spectral method  
to integrate the NLS equation which was 
specifically developed to handle rational initial data over unbounded domains
\cite{Islas2017}.
The main idea is to use Chebyshev points in conjunction with a map that
takes the interval $(0,\pi)$ to the infinite interval $(-\infty,\infty)$ \cite{boyd1987b}. The advantage of this approach is that  the spatial derivatives can be approximated using the Fast Fourier Transform (FFT) routine \cite{Trefethen} while the time evolution is obtained using a 4th-order Runge-Kutta integrator. This 4th-order in time Chebyshev pseudo-spectral scheme
(CPS4) was tested on the Peregrine solution and higher order rationals.
Using
 $N = 1024$ Chebyshev nodes and the time step $dT = 1.25\times 10^{-6}$, 
the  $H^1$-norm of the difference between the analytical and numerical solutions is at most ${\cal O}(10^{-7})$ for the test solutions.
Further, the 
global invariants,
the norm ${\cal I} = \int_{\cal D} |u|^2\,\mathrm{d}x$ 
and Hamiltonian 
$ {\cal H} =  -i\int_{\cal D} \left\{ | u_x |^2 - | u |^4 \right\}\,dx$,
are preserved to ${\cal O}(10^{-8})$.

To investigate the stability of the Peregrine solution, $u_P(x,t)$, we begin by considering random perturbations in the initial data typically encountered in experimental settings:
\be
P_\delta(x,0) = u_P(x,0) + \delta f_j(x), \qquad  j = 1, \dots,5,
\label{ic2}
\ee
where the amplitude of the background state is $a =0.5$, the  size of the perturbation is $\delta = 10^{-2}$  and
\begin{enumerate}
\item[a)] $f_1(x) = \cos2\pi (x+\phi)/L$ where $\phi\in [0,1]$ is a
random phase shift;
\item[b)] $f_2(x) = r(x)$ where $r(x)\in [0,1]$ is  random noise;
\item[c)] $f_3(x) = \sum_{k=1}^K\,r_k(x)\mbox{e}^{-(x-x_k)^2}$, representing $K$ localized Gaussian perturbations about $x_k$;
\item[d)] $f_4(x) = \sum_{k= low- modes}\,r_k(x)\mbox{e}^{i2\pi kx/L}$, 
representing a low frequency perturbation;
\item[e)] $f_5(x) = \sum_{k=high-modes}\,r_k(x)\mbox{e}^{i2\pi kx/L}$,
representing a high frequency perturbation.
\end{enumerate}
For each of the perturbations $f_j(x), j = 1, \dots,5$,  an ensemble of 100 numerical experiments, $ l = 1, \dots, 100$, was carried  out by varying the random components, $\phi$ or $r_k(x) \in [0,1]$, in 
the initial data.
The time frame of the experiments is 
$0 \le t \le 15$. 

Stability is determined by  monitoring the evolution of the difference of the perturbed solution and the nearest element of the unperturbed family.
If every perturbed solution $P_\delta(x,t)$ remains close (as defined below)
to its respective  nearest element of the family of Peregrine breathers, $u_P(x,t-t_l)$,  then this indicates the Peregrine breather is stable; otherwise the Peregrine solution is classified as unstable.

The  nearest element $u_P(x,t-t_l^*)$ to $P_\delta(x,t)$ is found as follows:
Let the difference of the perturbed, $P_\delta(x,t)$, and analytical, $u_P(x,t-t_l)$, solutions be  given by
\begin{align}
{\cal D}(t;t_l) &= ||P_\delta(x,t) - u_p(x,t-t_l)||_{H^1},\nonumber\\
\mbox{where}\quad
||u(x,t)||_{H^1} &= \int_{x_1}^{x_N} \left(|u_x|^2 + |u|^2\right)\,dx.
\end{align}
Then define
\begin{align}
{\cal D}_{max}(t_l) &= \max_{t\in[0,15]}{\cal D}(t;t_l),
\end{align}
and select the parameter value $t_l^*$ which minimizes
${\cal D}_{max}(t_l)$, i.e
\begin{align}
{\cal D}_{mm} &= \min_{t_l} {\cal D}_{max}(t_l) = 
{\cal D}_{max}(t_l^*).
\end{align}
As such,   the nearest element of the Peregrine family
is $u_P(x,t- t_l^*)$.
Further, whether  $P_\delta(x,t)$ remains close to $u_P(x,t- t_l^*)$
can be assessed using ${\cal D}_{mm}$.
For 
each $f_j(x)$ an  estimate of closeness  for the ensemble  is obtained using ${\cal A}_j(t)$,  the average of ${\cal D}_{mm}$ over all 100 simulations.

\begin{figure}[!ht]
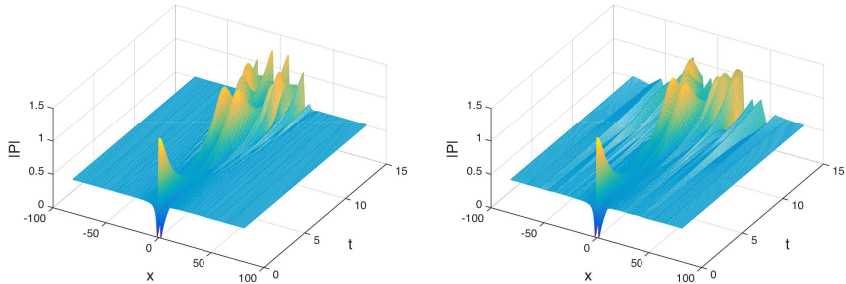

\centerline{
\includegraphics[width=0.47\textwidth]{growth_rate_SURF1_T0to15_011918}
\includegraphics[width=0.47\textwidth]{growth_rate_SURF2_T0to15_011918}
}
\caption{Sample surfaces $P_\delta(x,t)$ obtained using initial data (\ref{ic2}) with the (a) random phase $f_1(x)$ and (b) random noise $f_2(x)$ perturbations.}
\label{PertPer}
\end{figure}

\begin{figure}[!ht]
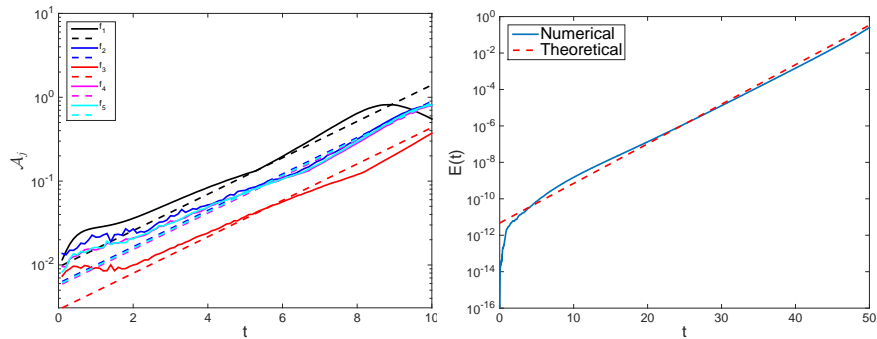

\centerline{
\includegraphics[width=0.47\textwidth]{growth_rate_PER_T0to10_011918}
\includegraphics[width=0.47\textwidth]{growth_rate_NPER_011918}
}
\caption{Growth of (a) ${\cal A}_j(t)$ for each perturbation type $f_j$ 
for $P_\delta$ experiments and  (b) Numerical error $E(t)$ for Peregrine initial data $u_P(x,0)$. In both plots the dashed lines represent the theoretically predicted maximal growth rate.}
\label{Growth}
\end{figure}

Figure~\ref{PertPer}  provides two  sample surfaces using  initial data
(\ref{ic2})  for (a)  the random phase perturbation $f_1$ and for (b) the random field perturbation $f_2$.
Clearly $P_\delta(x,t)$  does not remain close to the Peregrine  solution for these two perturbations.
In fact $P_\delta(x,t)$ doesn't remain close for any of the perturbations.
The semilog plot in Figure~\ref{Growth}(a) shows that  ${\cal A}_j(t)$ (shown with solid curves) grows exponentially fast  to ${\cal O}(1)$ for all the random perturbations $f_j(x)$  considered (the corresponding theoretical lines with the maximum growth rate are shown with dashed lines).
This exponential divergence of the solutions for various perturbed initial data indicates the Peregrine breather is linearly unstable.
It should be  noted that since the $f_j$ are not in the space
of polynomially decaying functions considered by the squared eigenfunction 
analysis, their growth rates are not predicted entirely by the analysis and   exhibit some variation from the maximum growth rate (pure noise $f_2$ is the most closely aligned).  Further numerical evidence  supporting the instability of the Peregrine solution can be found  in \cite{cuevas2017,kh2015}.

In the last experiment we examine the time evolution of the unperturbed Peregrine initial condition, $u_P(x,0)$, subject to numerical round off error. The obtained numerical solution, $U(x,t)$, can be viewed as a perturbation of the Peregrine solution and reflects the evolution of polynomially decaying perturbations in the solution space examined in the linear analysis of Section 4. Scheme CPS4 is prohibitively expensive to examine the growth of round off error. A more efficient scheme designed to handle polynomial decay is a pseudo spectral splitting scheme \cite{Islas2017}. We use $N = 8192$ fourier modes, $dT = 10^{-3}$ and $L = 800$.
The error is given by $E(t) = \max_x||u_P(x,t) - U(x,t)||$. Recall that when $a = 0.5$, $\sigma_{\max} = 2a^2 = 0.5$. The semilog plot in Figure~\ref{Growth}(b) shows that the error $E(t)$ grows at a rate that is in excellent agreement with the theoretical maximum growth rate $\sigma_{\max}$ (represented by the dashed line). 

In conclusion, our analysis shows the  Peregrine solution is linearly unstable, inheriting the instabilities of the underlying plane wave. 
Via the squared eigenfunctions we find exact exponentially growing solutions of the linearized NLS about $u_P(x,t)$ and  the theoretically predicted maximum growth rate $\sigma_{\max}$. 
The  numerical experiments confirm the squared eigenfunction analysis. Significantly, the numerically computed growth rates of the various perturbations were found to be in close agreement with the theoretical maximum growth rate, especially in the case of polynomially decaying perturbations.

\section*{Acknowledgements}
The authors gratefully acknowledge the support of the NSF and Simons Foundation: A. Calini through grant DMS-1109017 and C.M Schober through grant DMS-1108973 and Simons Award ID 527565. 

\section*{References}

%\bibliography{schober_APNUM}

\end{document}